\documentclass[aps,prl,twocolumn,superscriptaddress,longbibliography,floatfix]{revtex4-2}
\usepackage{amsmath,amssymb,mathrsfs,graphicx}
\usepackage{xcolor}
\usepackage{hyperref}
\hypersetup{colorlinks=true,allcolors=blue}

\begin{document}

\title{Geometrically Refined Chern Topology}

\author{Yu-Hao Wan}
\email{wanyh@stu.pku.edu.cn}
\affiliation{International Center for Quantum Materials, School of Physics, Peking University, Beijing 100871, China}
\author{Peng-Yi Liu}
\affiliation{International Center for Quantum Materials, School of Physics, Peking University, Beijing 100871, China}
\author{Qing-Feng Sun}
\email{sunqf@pku.edu.cn}
\affiliation{International Center for Quantum Materials, School of Physics, Peking University, Beijing 100871, China}
\affiliation{Hefei National Laboratory, Hefei 230088, China}

\begin{abstract}
Refined topological classifications usually rely on internal or crystalline symmetries. Here, we identify a symmetry-independent, geometry-constrained refinement of two-dimensional Chern topology for Bloch Hamiltonians. Collapse lines are closed momentum cycles whose Bloch-map images are self-retracing. These lines partition the Brillouin zone into independently compactifiable sectors and define an integer sector Chern vector. Within the collapse-partitioned Hamiltonian space, this vector distinguishes phases with the same total Chern number. Its change across a spatially uniform quench obeys a sector-resolved overlap-vortex index, enforcing total-time-refraction points with quantized net winding. Across a spatial interface, a sector Chern mismatch produces oppositely directed spectral flows when the conserved boundary momentum resolves the sectors. The construction extends to multiband occupied subspaces and provides a route to geometry-resolved topology without requiring a nontrivial internal or crystalline symmetry.
\end{abstract}

\maketitle
Topological photonics has emerged as a versatile platform for exploring exotic states of matter, enabling robust light transport immune to backscattering \cite{lu_topological_2014,ozawa_topological_2019,hafezi_imaging_2013,rechtsman_photonic_2013}. A central theme in this field is the classification of phases \cite{ryu_topological_2010}. At the coarse level, phases are distinguished by global topological invariants, such as the Chern number in quantum Hall systems \cite{thouless_quantized_1982,haldane_model_1988,wu_edge_2024}. At a finer level, the inclusion of symmetries, ranging from internal symmetries (such as time-reversal invariance) to spatial symmetries, allows for a richer classification of symmetry-protected topological phases \cite{Wieder2022,hafezi_robust_2011,qi_topological_2011,chiu_classification_2016,devescovi_axion_2024,wieder_axionic_2020,wan_classification_2025,Huang2025}. In these systems, phases that are indistinguishable by global invariants can be further resolved by symmetry-dependent topological indices, such as the spin Chern number and mirror Chern number, enabling unique phenomena like robust counter-propagating boundary channels and higher-order corner states \cite{khanikaev_photonic_2013,yin_photonic_2013,bliokh_quantum_2015,peterson_quantized_2018,xie_second-order_2018}.

It is widely accepted that once the protecting symmetries are broken, the fine-grained classification breaks down, leaving only the coarse global invariants to describe the system \cite{hasan_colloquium_2010,qi_topological_2011,chiu_classification_2016,fu_topological_2011,wen2012}. This leads to a question: Is the existence of refined topological states strictly bound to protecting symmetries? Or is it possible to distinguish phases with identical global invariants through a distinct fundamental mechanism that survives even when all such symmetries are removed?

In this work, we demonstrate that such a symmetry-free
refinement is possible through the geometry of the
Bloch-Hamiltonian map itself. Based on the generic Hamiltonian form $H(\mathbf{k}) = \mathbf{d}(\mathbf{k}) \cdot \boldsymbol{\sigma}$, we introduce a topological classification governed entirely by the geometric mapping of the $\mathbf{d}(\mathbf{k})$ vector. The core physical picture relies on the emergence of ``collapse lines''---singular contours in momentum space where the $\mathbf{d}$-vector mapping undergoes a dimensional collapse. These lines partition the Brillouin zone into distinct sectors, each of which independently maps to a closed geometry in $\mathbf{d}$-space. The Berry flux through a generic region of the Brillouin zone varies continuously under gapped deformations; the collapse lines make the flux through each sector an integer. This unique partitioning enables the definition of independent ``sector invariants'' that distinguish topological phases sharing the same total Chern number. We demonstrate that this geometry-refined topology dictates physical observables in both temporal and spatial domains: a time-boundary quench yields sector-resolved total-time-refraction points, while spatial interfaces host momentum-disjoint edge modes robust against long-range disorder. Symmetries refine topological classification through the transformation properties of Bloch states; collapse geometry refines it through the distribution of quantized Berry flux among momentum sectors. Our work provides a  geometric route to engineering robust dynamics and transport, broadening the scope of topological phases in photonic, acoustic, and topolectrical systems.
\begin{figure}
\begin{centering}
\includegraphics[width=\columnwidth]{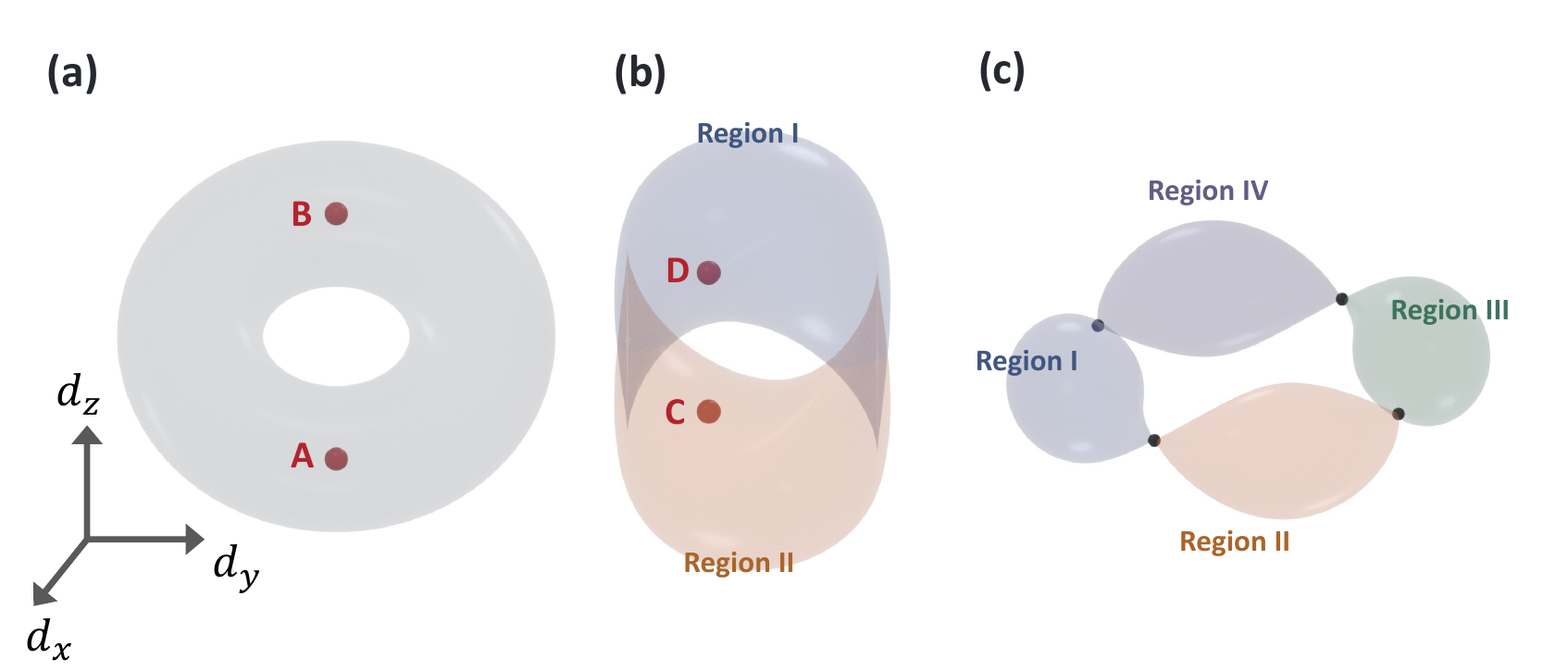
}
\par\end{centering}
\caption{\label{fig:1}
Schematic illustration of the mapping
$\mathbf d:T^2\rightarrow\mathbb R^3$ from the Brillouin zone to
$\mathbf d$ space.
(a) A connected $\mathbf d_{\mathbf k}$ surface bounding a single enclosed
region. A and B denote two possible positions of the origin within this same
region.
(b) A geometrically partitioned $\mathbf d_{\mathbf k}$ surface bounding two
separated enclosed regions. The origin lies in Region I for phase D and in Region II for phase C. (c) A multi-sector $\mathbf d_{\mathbf k}$ surface bounding four separated
enclosed regions, labeled I--IV.}
\end{figure}

\textit{Geometry-induced refined classification.}---We start from a generic two-band Hamiltonian
\[
H(\mathbf{k})=\mathbf{d}(\mathbf{k})\cdot\boldsymbol{\sigma} .
\]
The mapping $\mathbf{d}:\mathrm{BZ}\rightarrow\mathbb{R}^{3}$ provides
an intuitive geometric picture of band topology \cite{stone_mathematics_2009}. The entire image $\mathbf{d}(\mathbf{k}) \subset \mathbb{R}^3$ forms a closed surface, which we call the $\mathbf d_{\mathbf k}$ surface.
The relative position of the $\mathbf d_{\mathbf k}$ surface with respect to the origin encodes the global topology of the system. For instance, if the $\mathbf d_{\mathbf k}$ surface encloses the origin, the mapping is topologically nontrivial and yields a nonzero Chern number, which counts how many times $\mathbf{d}(\mathbf{k})$ wraps around the origin as $\mathbf{k}$ spans the Brillouin zone \cite{asboth_short_2016}. Our central observation is that the connectivity of the
$\mathbf d_{\mathbf k}$ surface provides a finer resolution of phases
sharing the same global winding.

We distinguish two generic mapping geometries, as illustrated in
Fig.~\ref{fig:1}. In Fig.~\ref{fig:1}(a), the
$\mathbf d_{\mathbf k}$ surface bounds a single enclosed region.
Configurations A and B correspond to two positions of the origin within this
same region. Both configurations have the same Chern number
$\mathcal C=1$. Because the relative position of the origin can be varied
continuously from A to B without crossing the $\mathbf d_{\mathbf k}$ surface,
the bulk gap remains open throughout the interpolation. Thus A and B are
adiabatically connected. In an explicit lattice model with this connected
geometry, the A$|$B interface carries no interface spectral flow, as shown
in Section S-7 of the Supplemental Material.

In contrast, the $\mathbf d_{\mathbf k}$ surface in
Fig.~\ref{fig:1}(b) bounds two separated closed regions, labeled I and II.
Phase D corresponds to the origin lying inside Region I, whereas phase C
corresponds to the origin lying inside Region II. In both cases, the surface encloses the origin once, and hence
the total Chern number remains $\mathcal C=1$. Nevertheless, as long as the
two-region mapping geometry is preserved, transferring the origin from one
region to the other requires it to cross the $\mathbf d_{\mathbf k}$ surface,
resulting in a bulk-gap closing. C and D therefore belong to distinct refined
topological sectors that cannot be distinguished by the total Chern number. More generally, the number of enclosed regions need not be limited to two. As illustrated in Fig.~\ref{fig:1}(c), an explicit Hamiltonian realization presented in Section S-3
of the Supplemental Material yields a $\mathbf{d}_{\mathbf{k}}$ surface with four distinct enclosed regions.
\begin{figure}
\begin{centering}
\includegraphics[width=\columnwidth]{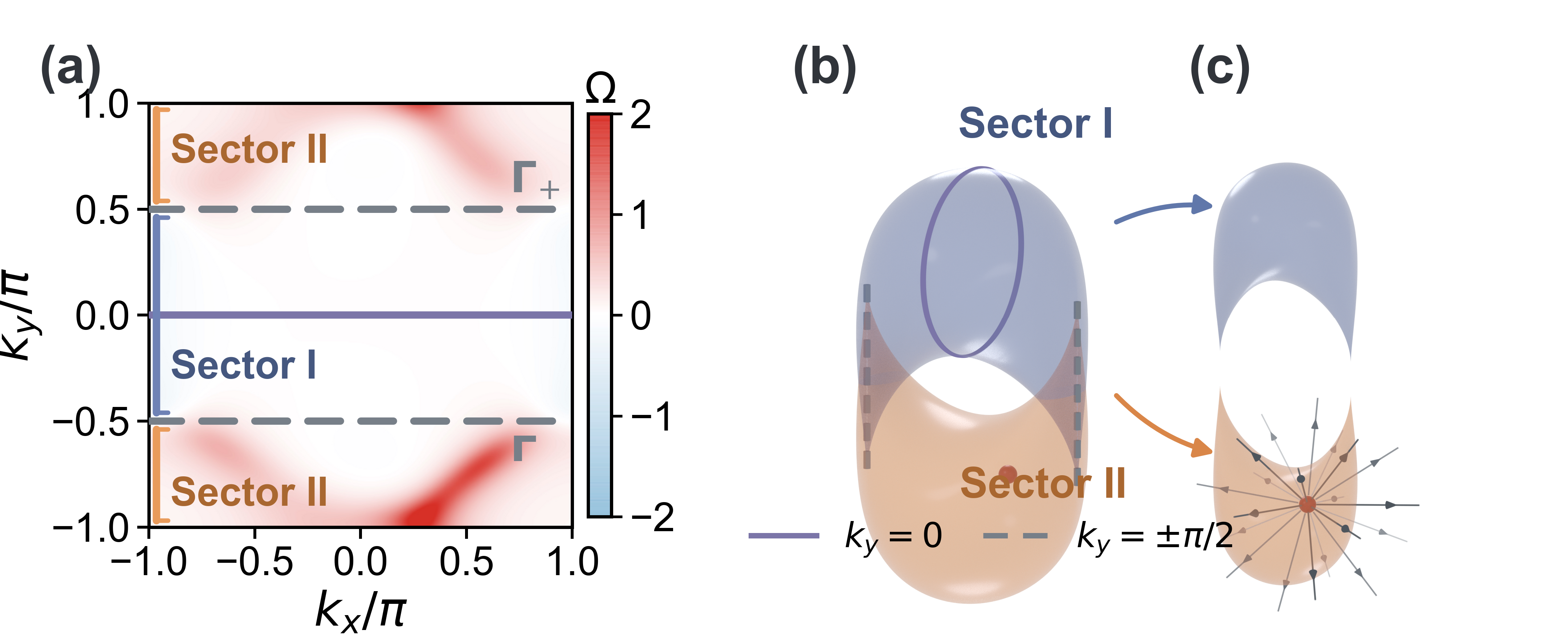}
\par\end{centering}
\caption{\label{fig:2}
(a) Berry curvature $\Omega_{xy}(\mathbf{k})$ for phase C
over the Brillouin zone,
with Sectors I and II and the contours
$k_y=0$ and $\Gamma_\pm:k_y=\pm\pi/2$ indicated.
(b) Image of the Brillouin zone in (a) under
$\mathbf d:\mathrm{BZ}\rightarrow\mathbb
{R}^{3}$.
The blue and orange sheets correspond to Sectors I and II.
(c) The two sector images displayed separately; the gray arrows indicate
the Berry flux.
}
\end{figure}

To understand the microscopic origin of this partitioned geometry and explicitly demonstrate such a refined classification, we consider the representative two-band Hamiltonian
\begin{equation}
\begin{aligned}
H_{\mathrm C/D}(\mathbf{k})={}&
(\sin k_x\cos k_y+\delta_x)\sigma_x\\
&+(\sin k_y+\delta_y)\sigma_y\\
&+(m+t_1\cos k_x+t_2\cos k_y)\sigma_z .
\end{aligned}
\label{eq}
\end{equation}
We take $t_1=1.5$, $t_2=2$, $\delta_x=0.3$, and $\delta_y=0.2$. The choices $m=+1.5$ and $m=-1.5$ define phases C and D, respectively, as illustrated schematically in Fig.~\ref{fig:1}(b). In both cases, the image of the Bloch Hamiltonian consists of two surface sectors enclosing distinct regions in $\mathbf d$ space. For these parameters, the model breaks time-reversal, particle-hole,
and global chiral symmetries, as well as the nontrivial unitary and
antiunitary square-lattice point-group symmetries. Structurally, this Hamiltonian is a finite-range anisotropic deformation of the Qi--Wu--Zhang model, which has been realized in a two-dimensional optical Raman lattice \cite{qi_topological_2006,liang_realization_2023}. Its symmetry analysis and relation to the Qi--Wu--Zhang model are provided in Sections S-1 and S-2 of the Supplemental Material. The microscopic origin of this partitioning lies in the geometric singularity of the mapping from the Brillouin zone to $\mathbf d$ space. To see this explicitly, we examine the evolution of constant-$k_y$ contours [Fig.~\ref{fig:2}(a)]. A generic slice, such as $k_y=0$, maps to a closed loop enclosing a finite area in $\mathbf d$ space [see purple curve in Fig.~\ref{fig:2}(b)]. By contrast, along the specific contours $k_y=\pm\pi/2$, the Hamiltonian reduces to
\begin{equation}
\mathbf d(k_x,\pm\pi/2)
=
\left(
\delta_x,
\delta_y\pm1,
m+t_1\cos k_x
\right).
\label{eq:collapse_line_image}
\end{equation}
Because $d_x$ and $d_y$ lose their $k_x$ dependence along these contours, the image of $k_x \in [-\pi, \pi]$ becomes self-retracing and collapses into a one-dimensional line segment with zero enclosed area. We therefore define these singular momentum-space contours, $\Gamma_\pm:k_y=\pm\pi/2$, as \emph{collapse lines}. As depicted in Fig.~\ref{fig:2}(a), these collapse lines naturally partition the Brillouin zone into Sector I ($|k_y|<\pi/2$) and Sector II ($\pi/2<|k_y|\leq\pi$). This momentum-space partitioning directly dictates the separated surface geometry in $\mathbf d$ space, with the sector images displayed together in Fig.~\ref{fig:2}(b) and separately in Fig.~\ref{fig:2}(c).


As shown in Fig.~\ref{fig:2}(c), the image of each momentum sector forms a closed surface in $\mathbf d$ space, with the collapse-line images serving as self-retracing seams. For a generic partition of the Brillouin zone, the Berry flux through an
individual momentum region is not quantized and may vary continuously
under a gapped deformation. The collapse geometry changes this situation:
the self-retracing collapse-line images allow each momentum sector to be
independently compactified without adding Berry flux. The Berry flux through each closed sector surface is therefore quantized, defining an independent sector Chern number,
\begin{equation}
\mathcal{C}_i
=
\frac{1}{2\pi}
\int_{\mathcal{I}_i}
\Omega_{xy}(\mathbf{k})\,\mathrm{d}^2k
\in\mathbb{Z},
\qquad
i=\mathrm{I},\mathrm{II}.
\label{eq:sector_chern}
\end{equation}
Here, $\Omega_{xy}(\mathbf{k})$ is the Berry curvature of the occupied band. Its momentum-space distribution, together with the collapse-sector partition, is shown in Fig.~\ref{fig:2}(a). We collect the sector Chern numbers into the vector
$\boldsymbol{\mathcal{C}}_{\mathrm{sec}}=(\mathcal{C}_{\mathrm{I}},\mathcal{C}_{\mathrm{II}})$,
whose entries sum to $\mathcal{C}_{\mathrm{tot}}$.
Direct integration of $\Omega_{xy}(\mathbf{k})$ over the two sectors yields
$\boldsymbol{\mathcal{C}}_{\mathrm{sec}}^{C}=(0,1)$ for phase C and
$\boldsymbol{\mathcal{C}}_{\mathrm{sec}}^{D}=(1,0)$ for phase D, although both have $\mathcal{C}_{\mathrm{tot}}=1$; the vector therefore identifies the sector carrying the quantized Berry flux.

For a labeled partition
$\mathcal P_N=\{\mathcal I_1,\ldots,\mathcal I_N\}$,
the gapped Hamiltonians are labeled by
$\boldsymbol{\mathcal C}_{\rm sec}\in\mathbb Z^N$,
with $\mathcal C_{\rm tot}=\sum_i\mathcal C_i$. The four-sector geometry illustrated in Fig.~\ref{fig:1}(c), for example, is characterized by a four-component sector Chern vector.

Note that this topological refinement is induced entirely by the collapse geometry and does not rely on any internal or spatial symmetries. We use the two-band representation because it provides the minimal setting in which the collapse sectors can be visualized directly in $\mathbf d$ space, while effective two-band descriptions are also common in synthetic lattices with two active modes. The construction is not restricted to two bands: for multiband systems, the sector Chern numbers are rigorously defined through the occupied-band projector, as formulated and demonstrated with an explicit four-band model in Section S-4 of the Supplemental Material. The sector Chern vector provides an integer-valued refinement
of the total Chern number whenever the collapse partition is present. As long as the bulk gap remains open and the collapse partition remains
well defined, the sector Chern vector is invariant under continuous
deformations. Therefore, a change of $\boldsymbol{\mathcal C}_{\rm sec}$ 
necessarily requires the bulk gap to close and reopen.
We next show that this refinement has a direct bulk dynamical consequence.
\begin{figure}[htbp]
\centering
\includegraphics[width=1\linewidth]{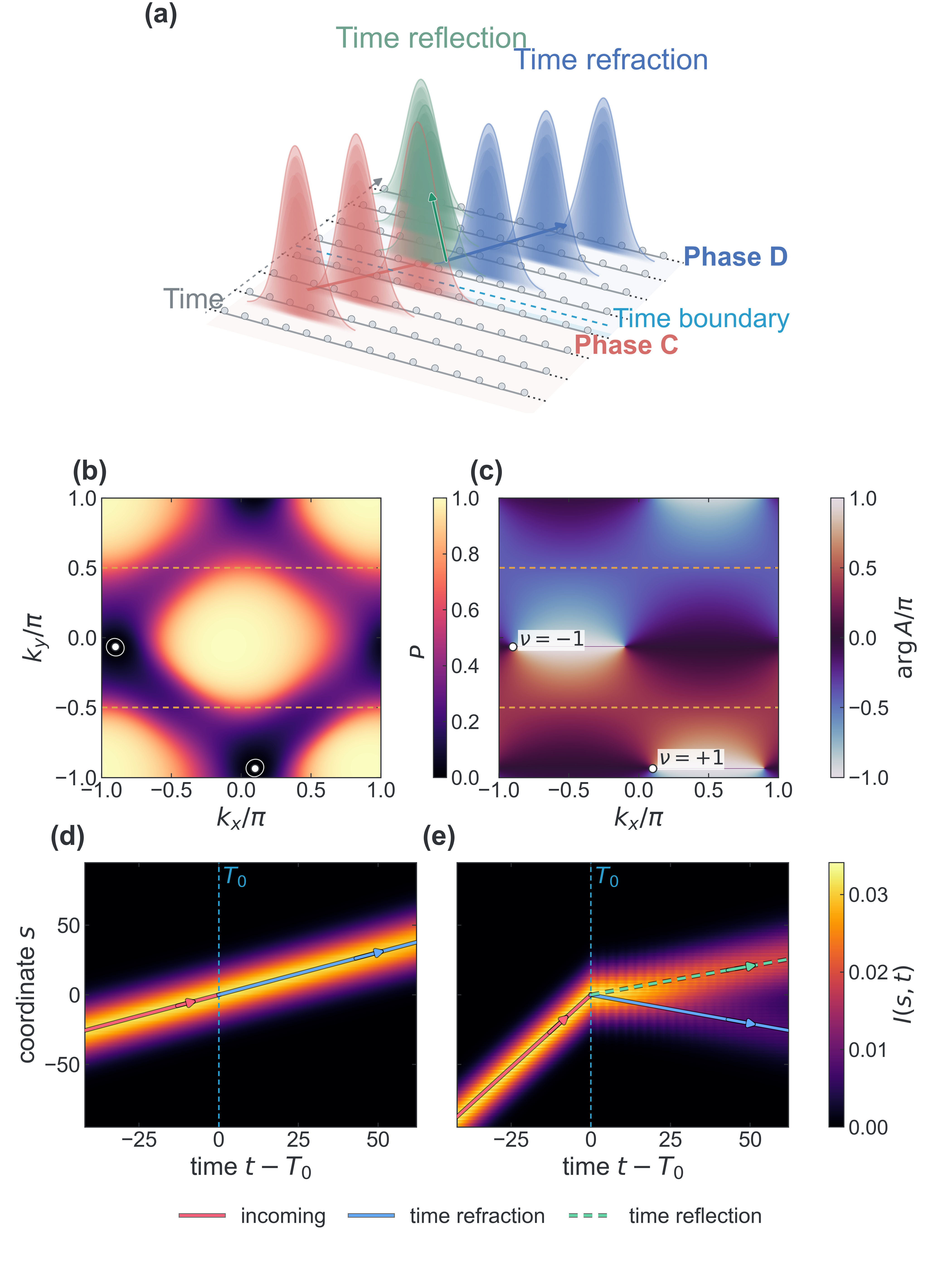}
\caption{
{Bulk time-boundary signature of sector Chern transfer.}
(a) Schematic of the spatially uniform $C\rightarrow D$ quench.
(b) Post-quench lower-band population
$P_{--}(\mathbf k)=|\langle u_D^-(\mathbf k)|u_C^-(\mathbf k)\rangle|^2$.
White circles mark the total-time-refraction points, and dashed lines mark the
collapse lines.
(c) Phase of the overlap amplitude
$A(\mathbf k)=\langle u_D^-(\mathbf k)|u_C^-(\mathbf k)\rangle$.
The two zeros form vortices with opposite winding numbers in the two collapse sectors.
(d,e) Projected wave-packet dynamics for packets centered at
$\mathbf k_0/\pi=(-0.901,-0.064)$ and
$\mathbf k_c/\pi=(0.4,-0.3)$, respectively.
Only the time-refracted branch remains at $\mathbf k_0$, whereas the generic
packet splits into two temporal branches. Vertical dashed lines mark the quench
time $T_0$, and guide lines indicate the post-quench branch trajectories.
}
\label{fig:3}
\end{figure}

\textit{Bulk time-boundary probe of sector topology.}---To directly detect the refined sector topology in the bulk, we employ a spatially uniform time-boundary quench. Because such a global quench strictly preserves the crystal momentum $\mathbf k$, it serves as a natural momentum-resolved dynamical probe. An incident Bloch state at a fixed $\mathbf k$ is projected onto the two post-quench temporal branches [Fig.~\ref{fig:3}(a)] \cite{moussa_observation_2023,ren_observation_2025}.

For the $C\rightarrow D$ quench, the Hamiltonian is switched from $H_C(\mathbf k)$ to $H_D(\mathbf k)$ at $t=T_0$. We denote the normalized Bloch eigenstates by $|u_{C/D}^\alpha(\mathbf k)\rangle$, where $\alpha=-,+$ labels the lower and upper bands. An initial lower-band state $|u_C^-(\mathbf k)\rangle$ populates the two post-quench branches of phase D with probabilities
\begin{equation}
P_{\alpha-}(\mathbf k)
=
\left|
\langle u_D^\alpha(\mathbf k)|u_C^-(\mathbf k)\rangle
\right|^2,
\qquad \alpha=\pm .
\end{equation}

Figure~\ref{fig:3}(b) shows the calculated lower-branch population $P_{--}(\mathbf k)$ after the quench. Strikingly, two isolated zeros emerge, situated precisely within Sector I and Sector II. At these specific momenta, $P_{--}=0$ and $P_{+-}=1$; the time-reflected channel is completely suppressed, leaving only the time-refracted branch \cite{li_observation_2026}. We refer to these momentum-space singularities as \emph{total-time-refraction points}.

The topological origin of these points is encoded in the phase of the complex overlap amplitude $A(\mathbf k)=\langle u_D^-(\mathbf k)|u_C^-(\mathbf k)\rangle$, for which $P_{--}=|A|^2$. As visualized in Fig.~\ref{fig:3}(c), each zero forms the core of an overlap-amplitude vortex. Its winding number $\nu_j$ and the corresponding sector sum satisfy
\begin{equation}
\begin{aligned}
\nu_j
&=
-\frac{1}{2\pi}
\oint_{\ell_j}
\nabla_{\mathbf k}\arg A(\mathbf k)\cdot d\mathbf k,\\
\sum_{\mathbf k_j\in\mathcal I_a}\nu_j
&=
\mathcal C_a^C-\mathcal C_a^D .
\end{aligned}
\end{equation}
Here, $\ell_j$ represents a closed loop enclosing the $j$th zero. Using Stokes' theorem, we prove in Section S-5 of the Supplemental Material that this
relation holds generally whenever the initial and final Hamiltonians share
the same collapse-sector partition. A nonzero sector Chern transfer therefore necessarily enforces one or more
total-time-refraction points within that sector, with their net vortex
winding fixed by the difference between the initial and final sector Chern
numbers. For the present quench, $\boldsymbol{\mathcal C}_{\mathrm{sec}}^C - \boldsymbol{\mathcal C}_{\mathrm{sec}}^D = (-1,1)$, which is in precise agreement with the opposite vortex winding numbers observed in Fig.~\ref{fig:3}(c). Their total winding over the entire Brillouin zone vanishes, consistent with the identical total Chern numbers ($\mathcal C_{\mathrm{tot}}=1$), yet their opposite sector-resolved windings directly reveal the
sector-Chern difference that is invisible to the total Chern number.

To visualize this momentum-space signature dynamically in real space, we prepare a finite-width Gaussian wave packet in the lower band of phase C, centered at an incident momentum $\mathbf k$, and switch the Hamiltonian globally to phase D at $t=T_0$. We plot the projected density $I(s,t)=\int ds_\perp\,\rho(s,s_\perp,t)$, where $s$ is the coordinate along the incident group-velocity direction and $s_\perp$ is the transverse coordinate. For $\mathbf k=\mathbf k_0$, corresponding to a total-time-refraction point
enforced by the sector Chern transfer, only the time-refracted branch survives the quench [Fig.~\ref{fig:3}(d)]. By contrast, a wave packet centered at a generic momentum $\mathbf k_c$ inevitably splits into both time-reflected and time-refracted branches [Fig.~\ref{fig:3}(e)]. The time-boundary response therefore effectively converts the abstract sector Chern transfer into a striking, observable bulk dynamical signature. We note that the required state preparation, abrupt Hamiltonian quench, and time-resolved
wave-packet detection have already been demonstrated in atomic momentum lattices, where time reflection and refraction were directly observed \cite{dong_quantum_2024}. Furthermore, complete conversion into a single post-quench branch has been successfully resolved through momentum-dependent temporal-scattering measurements in programmable circuit lattices \cite{li_observation_2026}.
\textit{Sector-resolved interface spectral flow.}---Complementary to the temporal bulk probe, the sector Chern mismatch has a
direct spatial manifestation whenever the conserved boundary momentum
resolves the distinct collapse sectors. For a boundary normal to the $x$ direction (where $k_y$ remains conserved), the total Chern number mismatch vanishes ($\Delta\mathcal C_{\mathrm{tot}}=0$). However, the sector Chern mismatch $\Delta\boldsymbol{\mathcal C}_{\mathrm{sec}} = \boldsymbol{\mathcal C}_{\mathrm{sec}}^{C} - \boldsymbol{\mathcal C}_{\mathrm{sec}}^{D} = (-1,1)$ requires oppositely directed spectral flows in the two momentum sectors.

\begin{figure}
\begin{centering}
\includegraphics[width=\columnwidth]{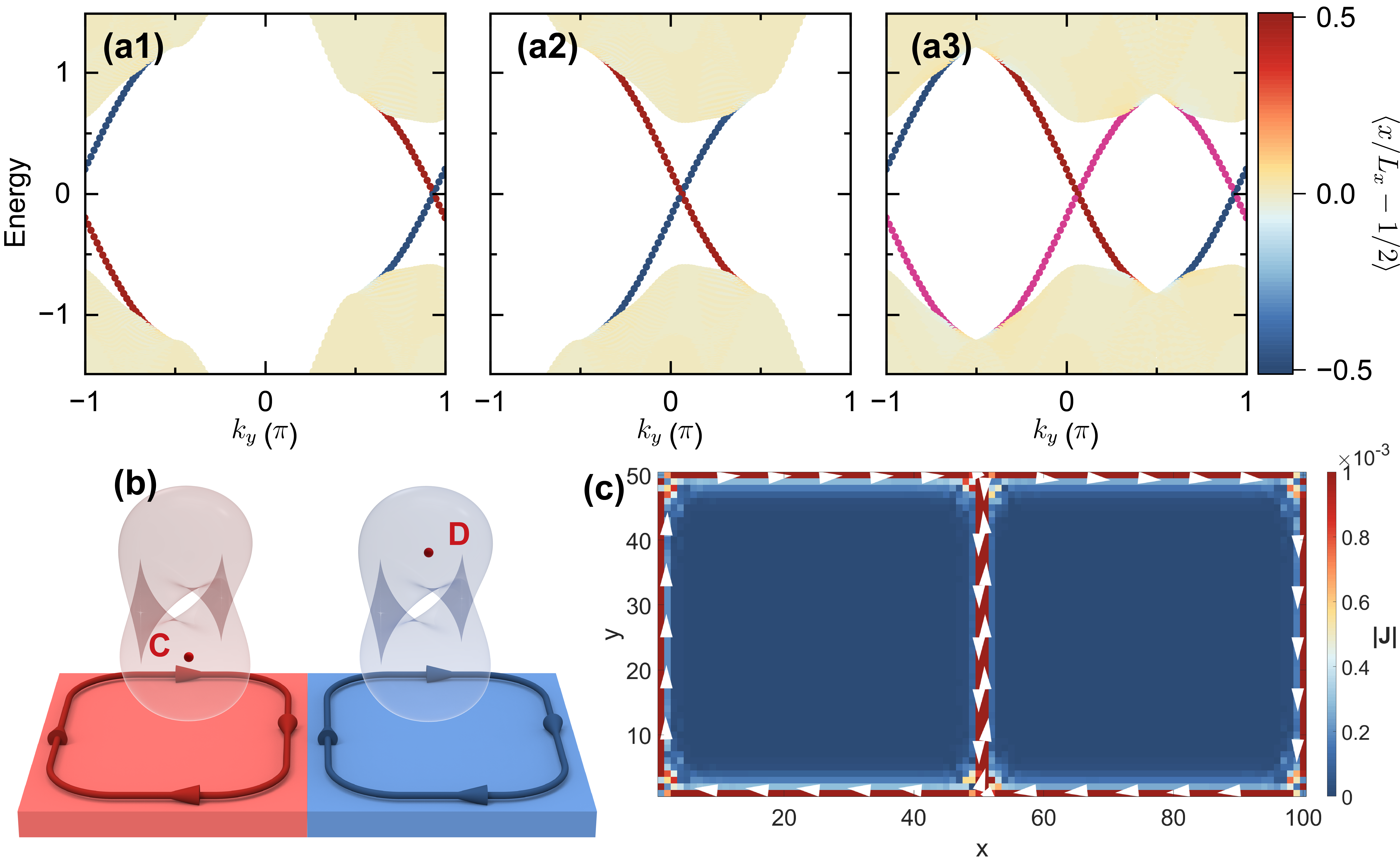}
\par\end{centering}
\caption{\label{fig:4}
(a1)--(a3) Ribbon spectra of phases C, D, and the C$|$D
heterostructure, respectively. In these panels, the color scale encodes the
mean transverse position $\langle x/L_x-1/2\rangle$. The pink overlay in
(a3) highlights the interface-localized spectral branches as a guide to the
eye and is independent of the position color scale.
(b) Real-space schematic of the C$|$D heterostructure; the red and blue
domains denote phases C and D, respectively.
(c) Equilibrium-current distribution at $E=0.2$. The color scale gives the
current magnitude $|J|$, and the arrows indicate the current direction.
Here, $L_x=100$ and $L_y=50$, with phase C for $x\leq50$ and phase D for
$x>50$.}
\end{figure}

Figures~\ref{fig:4}(a1) and \ref{fig:4}(a2) show the ribbon spectra of phases C and D, respectively, calculated under open boundary conditions along $x$ ($L_x=100$) while preserving translational symmetry along $y$. The color encodes the average transverse displacement $\langle x/L_x - 1/2 \rangle$ for each Bloch state, clearly distinguishing the boundary-localized branches. While both spectra display chiral edge states consistent with their identical total Chern number ($\mathcal C_{\mathrm{tot}}=1$), their boundary branches emerge in complementary $k_y$ intervals separated by the collapse lines at $k_y=\pm\pi/2$. When the two phases are joined to form a heterostructure ribbon ($x \leq 50$ for phase C and $x > 50$ for phase D), these branches form momentum-disjoint, counter-propagating channels at the internal C$|$D interface that directly bridge the bulk valence and conduction bands [Fig.~\ref{fig:4}(a3)]. The corresponding real-space equilibrium current distribution confirms that these oppositely directed channels are localized at the internal junction [Figs.~\ref{fig:4}(b,c)]. The detailed calculation of the equilibrium current distribution is provided in Section S-6 of the Supplemental Material.

This boundary response depends on interface orientation. For the
$x$-normal interface, the conserved momentum $k_y$ keeps the two spectral
flows separated in momentum space and therefore forbids their direct
hybridization. If the interface is instead oriented so that $k_x$ is
conserved, the projections of the two momentum sectors overlap and the
counter-propagating modes can hybridize and reconnect. Detailed calculations
of this orientation dependence and long-range-disorder response are
provided in Sections S-9 and S-8 of the Supplemental Material, which include
Refs.~\cite{buttiker_absence_1988,fisher_relation_1981,
wan_quantum_2025,kawarabayashi_quantum_2009,wan_interplay_2025}.

We have shown that collapse geometry enables a geometrically refined classification of Chern phases. Whereas the total Chern number records the net wrapping of the Bloch-Hamiltonian map, collapse lines resolve how the quantized Berry flux is distributed among independently compactifiable sectors. The resulting sector Chern vector remains invariant under gap-preserving deformations that maintain the labeled collapse partition. Across a time boundary, a change in the sector Chern vector enforces overlap-amplitude vortices, whose cores dynamically appear as total-time-refraction points, while a sector Chern mismatch across a spatial interface produces momentum-disjoint counter-propagating modes with robustness against long-range disorder.

More broadly, these results show that topological information beyond a global invariant can be encoded directly in the connectivity of the Bloch-Hamiltonian map. Topological classification may therefore depend not only on how many times a Bloch-Hamiltonian map wraps its target space, but also on where the quantized wrapping is stored. This geometry-based viewpoint also motivates extensions to more general
collapse partitions in higher-dimensional systems and corresponding geometry-resolved responses in photonic, acoustic, atomic, and circuit platforms. Collapse geometry thus provides a symmetry-free route to refined topology beyond the total Chern number.

\begin{acknowledgments}
We acknowledge the High-performance Computing Platform of Peking University for providing computational resources. This work was supported by the National Key Research and Development Program of China [2024YFA1409002]; the National Natural Science Foundation of China [12374034 and 124B2069]; and the Quantum Science and Technology-National Science and Technology Major Project [2021ZD0302403].
\end{acknowledgments}

\bibliographystyle{apsrev4-2}
\bibliography{ref}

\end{document}